\documentclass[aps,prb,preprint]{revtex4}
\usepackage{color}
\usepackage{amssymb}
\usepackage{dcolumn}
\usepackage{amstext}
\usepackage{graphicx}
\usepackage{amsmath}
\usepackage{hyperref}
\usepackage{amsthm}
\usepackage{lscape}
\usepackage{longtable}
\usepackage{amsfonts}
\usepackage{amssymb}
\usepackage{subfigure}
\usepackage{latexsym}

\newcommand{\be}{\begin{equation}}
\newcommand{\ee}{\end{equation}}
\newcommand{\ba}{\begin{eqnarray}}
\newcommand{\ea}{\end{eqnarray}}

\newcommand{\etal}{{\it et al}}

\begin{document}

\title{Effect of the pseudogap on T$_c$ in the cuprates and implications for its origin}
\author{Vivek Mishra$^1$}
\author{U. Chatterjee$^2$}
\author{J. C. Campuzano$^{1,3}$}
\author{M. R. Norman$^{1*}$}
\affiliation{$^1$Materials Science Division, Argonne National Laboratory, Argonne, IL 60439\\
$^2$Department of Physics, University of Virginia, Charlottesville, VA 22904\\
$^3$Department of Physics, University of Illinois, Chicago, IL 60607}

\date{\today}
\linespread{1}
\begin{abstract}
One of the most intriguing aspects of cuprates is a large pseudogap 
coexisting with a high superconducting transition temperature.
Here, we study pairing in the cuprates from electron-electron interactions
by constructing the pair vertex using spectral functions
derived from angle resolved photoemission data for a near optimal 
doped Bi$_2$Sr$_2$CaCu$_2$O$_{8+\delta}$
sample that has a pronounced pseudogap.
Assuming that that the pseudogap is {\it not} due to pairing,
we find that the
superconducting instability is strongly suppressed, in stark contrast
to what is actually observed.  Using an analytic approximation for the 
spectral functions, we can trace
this suppression to the destruction of the BCS logarithmic singularity from
a combination of the pseudogap and lifetime broadening.
Our findings strongly support those theories of the cuprates where
the pseudogap is instead due to pairing.
\end{abstract}

\maketitle
\linespread{1}
The origin of high temperature superconductivity remains one of
the most intriguing problems in physics.  A particularly dramatic
observation in the cuprates is the presence of a large pseudogap that 
spans much of the doping-temperature phase diagram \cite{Timusk,NPK}.  The origin
of this pseudogap is as debated as the mechanism for
superconductivity.  In one class of theories, the pseudogap arises
from some instability not related to pairing, typically charge,
spin, or orbital current ordering.  Recent evidence for this has come
from a variety of measurements indicating symmetry breaking \cite{Kaminski,Bourges,Kapitulnik,Shekhter}.
On the other side is a class of theories where the pseudogap is
associated with pairing.  This ranges from preformed pairs \cite{Emery} to
strong coupling `RVB' theories where singlet spin pairs become
charge coherent \cite{Wen}.  To date, numerical simulations, even on simple
models such as the single band Hubbard and t-J models, have yet
to definitively rule out one class in favor of the other.

For conventional superconductors, the development of the strong
coupling Eliashberg approach \cite{SSW} led to a detailed proof that the
electron-phonon interaction was the cause of superconductivity.
These equations could be inverted to derive the spectral function
of the bosons responsible for the pairing, and this spectrum matched
the phonon spectrum observed in the material \cite{McMillan}.  Such methods have
been employed as well in cuprates \cite{Huang}, but remain controversial since
the strong momentum dependence of the interaction (responsible
for the d-wave pairing) invalidates the approach used by McMillan
and Rowell \cite{McMillan} based on tunneling data, which by definition are
momentum averaged.  This is in turn greatly
complicated by the strongly anisotropic pseudogap, since the
 underlying assumption is a gapless normal state with weak momentum 
 dependence.
 Here, we address this issue by directly using angle resolved photoemission
(ARPES) data to construct the pair vertex.  Under the assumption that the
pseudogap is unrelated to pairing, we find no superconducting
instability.  We trace this effect to the suppression of the BCS logarithmic
instability by the pseudogap itself.

To construct the pair vertex, we must first know the single particle
Green's function.  An issue is that ARPES only measures occupied
states.  As in earlier work, we surmount this difficulty under the assumption
of particle-hole symmetry with respect to the Fermi energy and Fermi surface ($k_F$)
\begin{equation}
 A(k,\omega)=I(k,\omega)+I(-k+2k_F,-\omega)
 \label{eq:Akw_condition}
\end{equation}
where $I$ is the photoemission intensity and $\omega$ is measured relative to the
chemical potential.  Relaxing this approximation should only lead to minor differences
in the results.  The assumption that the left hand side of
the equation can be equated to the spectral function (imaginary part of the single
particle Green's function) requires subtracting any background from the intensity
(obtained from data for unoccupied momenta well beyond $k_F$), and 
then normalizing by requiring
the integrated weight over frequency to be equal to unity.  A similar method has
been successfully employed by us in several works, most recently \cite{Utpal07} to construct
the dynamic susceptibility in cuprates, which was found to be in good agreement with inelastic neutron
scattering (INS) data.  In fact, the data set we employ here, from a near optimal doped Bi$_2$Sr$_2$CaCu$_2$O$_{8+\delta}$
sample with a $T_c$ of 90 K, was used in that work to reproduce the momentum
and energy dependence of the INS data in the superconducting state, in particular
the unique hourglass-like dispersion observed in a variety of cuprates.  In our
case, though, we will use normal state data above T$_c$.  For this sample, a
relatively complete momentum sweep was done in an octant of the Brillouin
zone at a temperature of 140 K \cite{Kaminski01}, with data obtained on a 2 meV energy grid down to
322 meV below the Fermi energy, with the background intensity adjusted to match
each spectrum at this lower energy cut-off.

To proceed, we will assume that pairing originates from electron-electron interactions.
Although the particular approach used here is based on spin fluctuations, we believe
the results are general to any electronic pairing mechanism.  This first requires
constructing the polarization bubble
\begin{equation}
 \chi_0(q,\Omega) = \int_{-\infty}^{\infty}d\omega \int_{-\infty}^{\infty} d\omega^{\prime} 
 \frac{f(\omega)-f(\omega^\prime)}{\omega-\omega^\prime+\Omega+i0^{+}}
 \frac{1}{N}\sum_{k}A(k+q,\omega)A(k,\omega^{\prime})
\end{equation}
with $f(\omega)$ the Fermi function and $N$ the number of $k$ points.  The $k$ sum is
restricted to two dimensions under the further assumption of a single band (for the data set
we use, there is no evidence for bilayer splitting).  We will
then make the standard random phase approximation to construct the full dynamic
susceptibility
\begin{equation}
 \chi(k,\Omega) = \frac{\chi_0(k,\Omega)}{1-U \chi_0(k,\Omega)}
  \label{eq:chi}
 \end{equation}
 where $U$ is an effective screened Hubbard interaction appropriate for a single band involving
 hybridized copper 3d and oxygen 2p orbitals.
 
 In Fig.~1, we show the imaginary
 part of $\chi$ for two different values of $U$ along the $(0,0)-(\pi,\pi)$ direction.
 For the larger value of $U$ (860 meV), Im $\chi$ is concentrated at low frequencies
 at the commensurate wavevector $(\pi,\pi)$.  This is typical of very underdoped
 samples near the commensurate antiferromagnetic phase \cite{Stock}.  We contrast this
 with a smaller value of $U$ (800 meV), where spectral weight is now concentrated at
 incommensurate wavevectors at a higher energy, being a more appropriate description of
 INS data \cite{Hinkov} for slightly underdoped samples (consistent with the ARPES data set employed).
 Decreasing $U$ even further reduces the magnitude of Im $\chi$, moves the
 incommensurate weight to even higher energies, and suppresses the lower energy commensurate
 weight.

Using this $\chi$, the resulting electron-electron interaction is \cite{Monthoux,Scalapino}
\begin{equation}
 V(k,\Omega) = \bar{U}^{2} \left[\frac{3}{2} \chi(k,\Omega) -\frac{1}{2}\chi_0(k,\Omega)\right]
 \label{eq:Pairing_potential}
\end{equation}
where $\bar{U}$ differs from $U$ because of vertex corrections \cite{Vilk}.  To set $\bar{U}$, we will require
that the renormalized Fermi velocity at the d-wave node (the Fermi surface along the $(0,0)-(\pi,\pi)$ 
direction) matches that determined from the ARPES dispersion (1.6 eV$\AA$) assuming
a bare velocity of 3 eV$\AA$ from band theory.  The renormalization factor (3/1.6) can be
obtained as
\begin{equation}
Z = \left[1-\frac{\partial \Sigma^\prime}{\partial \omega}\right]_{\omega=0}
\end{equation}
where $\Sigma^\prime$ is the real part of the fermion self-energy, and
we assume $Z$ arises from the same interaction $V$ as above:
\begin{equation}
 \Sigma(k,i\omega_n)=T\sum_{q,\omega_m} V(k-q,i\omega_n-i\omega_m)G_0(q,i\omega_m)
 \label{eq:self_energy}
\end{equation}
where $G_0$ is the bare fermion Green's function
\begin{equation}
 G_0^{-1}(k,i\omega_n)=i\omega_n-\xi_k
 \label{eq:bare_g}
\end{equation}
and $\xi_k$ is the bare dispersion (obtained from a tight binding fit to the ARPES dispersion by multiplying by the renormalization
factor 3/1.6 mentioned above).  The real part of $\Sigma$ is then obtained by analytic continuation.
For the case shown in Fig.~1a, $\bar{U}$ turns out to be the same as $U$.  But for the case shown in Fig.~1b,
we must increase $\bar{U}$ to 928 meV to obtain the same $Z$.

We now turn to the pairing problem.  The anomalous (pairing) self-energy in the singlet channel is \cite{Monthoux,Scalapino}
\begin{equation}
 -\frac{T}{N}\sum_{k^\prime,\omega_m} V(k-k^\prime,i\omega_n-i\omega_m) 
  \mathcal{P}_0(k^\prime,i\omega_m)\Phi(k^\prime,i\omega_m) = \Phi(k,i\omega_n)
 \label{eq:gap_eq}
\end{equation}
with the pairing kernel $\mathcal{P}_0$
\begin{equation}
 \mathcal{P}_0(k^\prime,i\omega_m)=G(k^\prime,i\omega_m)G(-k^\prime,-i\omega_m). 
 \label{eq:p0}
\end{equation}
It is numerically convenient to solve this `linearized' gap equation in the
Matsubara representation (see Supplementary Information)
\begin{equation}
 \mathcal{F}(k,i\omega) = -\int^{+\infty}_{-\infty}\frac{dx}{\pi} \frac{\mathcal{F}^{\prime\prime}(k,x)}{i\omega-x},
 \label{eq:spectral_representation}
\end{equation}
where $\omega$ is the bosonic (fermionic) Matsubara frequency, for a given bosonic (fermionic)
function $\mathcal{F}$.
In Eq.~(\ref{eq:p0}), $G$ is the fully dressed Green's function, which is formally determined 
by including the self-energy correction Eq.~(\ref{eq:self_energy}) in a completely self-consistent approach.
Instead, we obtain $G$ from the experimental spectral functions as discussed above using Eq.~(\ref{eq:spectral_representation}).
This is related to the approach of Dahm {\it et al} \cite{Dahm} where INS data were used instead.

At T$_c$, the maximum eigenvalue ($\lambda_{max}$) of Eq.~(\ref{eq:gap_eq}) reaches unity and the corresponding
eigenvector gives the energy-momentum structure of the superconducting order parameter.
Ideally, we would need to know $G$ at each temperature.  This is impractical when using real experimental
data.  Instead, we use our experimental normal state data at 140 K, and assume that all temperature dependence
arises from the Matsubara frequencies.  As we will see below, this is a best case scenario, since if anything,
the magnitude of the pseudogap increases as the temperature is lowered.

Our results are shown in Fig.~2, labeled as FBZ (full Brillouin zone).  We see that $\lambda_{max}$ (which occurs for 
B$_{1g}$, i.e., d-wave, symmetry) is much less than unity and essentially temperature
independent for both cases shown in Fig.~1.  This implies that there is no superconductivity.  This is the central
result of our paper.

To understand this surprising result, we now turn to Fermi surface restricted calculations, which is a commonly employed
approximation where the momentum perpendicular to the Fermi surface ($k_{\perp}$) is integrated out.
This approximation is equivalent to ignoring the dependence of $V$ on $k_{\perp}$.
This procedure results in an equation which depends only on the angular variation around the Fermi surface:
 \begin{equation}
 -\frac{T}{N_\phi}\sum_{\phi^{\prime},\omega_m}  V^{\phi \phi^{\prime}}_{nm}\mathcal{P}_0(\phi^{\prime}, i\omega_m)
 \Phi(\phi^{\prime},i\omega_m) = \Phi(\phi,i\omega_n)
 \label{eq:Phi_fsr}
\end{equation}
where $N_\phi$ is the number of angular points and $V^{\phi \phi^{\prime}}_{nm}$ is
\begin{equation}
V^{\phi \phi^{\prime}}_{nm}=V(k_{Fx}^{\phi}-k_{Fx}^{\phi^{\prime}},k_{Fy}^{\phi}-k_{Fy}^{\phi^{\prime}},i\omega_n-i\omega_m).
\end{equation}
$\mathcal{P}_0$ is obtained by numerically integrating Eq.~(\ref{eq:p0}) using the experimental $G$ over $k_{\perp}$,
with the integration direction for each angle $\phi$ determined from the normal given by the tight binding fit to the ARPES data
(this same procedure is used to identify $k_F$ in Eq.~(\ref{eq:Akw_condition})).

The results are also shown in Fig.~2.  Although $\lambda_{max}$ is now temperature dependent, over the temperature range shown,
it is still below unity.  Paradoxically, $\lambda_{max}$ increases with increasing temperature.  We have verified that
at even higher temperatures, $\lambda_{max}$ reaches a maximum, and then begins to fall, with the more realistic second case
(Fig.~1b) always remaining below unity.  Similar behavior for the $T$ dependence of $\lambda$ was reported by Maier {\it et al} \cite{Maier}.

To understand this behavior, we now turn to some analytic calculations.  To a good approximation, we can approximate $V$ for the
d-wave case in the weak coupling BCS limit as
\begin{equation}
 V(\phi,\phi^\prime)= \mathcal{V} \cos(2\phi) \cos(2\phi^\prime)
\end{equation}
and assume an isotropic density of states $N_0$ over the Fermi surface coming from $\xi_k$. The weak coupling equation for $T_c$ is
\begin{equation}
T\sum_{\omega_n} \int_{0}^{2\pi}\frac{d\phi}{2\pi} \mathcal{V} \cos^2(2\phi) P_0(\phi,i\omega_n) =1.
\label{eq:Tcweak}
\end{equation}
For $G$ we use a phenomenological form that is a good representation of ARPES data \cite{Norm07}
\begin{equation}
 G(k,i\omega_n) = -\frac{i\omega_n+ i\Gamma sgn(\omega_n)+\xi_k}{(\omega_n+\Gamma sgn(\omega_n))^2+\xi_k^2+\Delta^2_k}.
 \label{eq:model_G}
\end{equation}
Here $\Gamma$ is the broadening and $\Delta_k$ the anisotropic pseudogap, which consistent with ARPES, is assumed to have a d-wave anisotropy.
On the Fermi surface, this can be approximated as $\Delta_0 \cos(2\phi)$. The pairing kernel can now be analytically
derived
\begin{equation}
 \mathcal{P}_0(\phi,i\omega_n)=\pi N_0 \left[\frac{1}{\sqrt{\tilde{\omega}^2_n+\Delta^2_{\phi}}}-\frac{\Delta^2_{\phi}}{2(\tilde{\omega}^2_n+\Delta^2_{\phi})^{3/2}}\right].
 \label{eq:pair_kernel_pg}
\end{equation}
Here $\tilde{\omega}_n$ is $\omega_n+sgn(\omega_n)\Gamma$. To obtain an analytic approximation,
we replace the sum $T\sum_{\omega_n}$ by an integral $\int d\omega/2\pi$, using the Euler-Maclaurin formula \cite{Abramowitz}
for low temperatures in Eq.~(\ref{eq:Tcweak}) and
rewrite the condition for T$_c$ as
\begin{equation}
1 = N_0 \mathcal{V} \int_{\pi T}^{\infty} d\omega \int^{2\pi}_{0} \frac{d\phi}{2\pi} \cos^2 2\phi \left[\frac{1}{\sqrt{\tilde{\omega}^2+\Delta^2_{\phi}}} \right.
- \left. \frac{\Delta^2_{\phi}}{2(\tilde{\omega}^2+\Delta^2_{\phi})^{3/2}}\right].
\label{pot3} 
\end{equation}
The integral over $\omega$ can be carried out analytically. The second term 
is convergent, so we can integrate it to $\infty$.
For the first term, 
we use a BCS cut-off energy $\omega_c$ and
we assume $\omega_c \gg T, \Delta_0, \Gamma$ and in the low $T$ limit we get
\begin{equation}
1 \simeq N_0 \mathcal{V}  \int_{0}^{2\pi}\frac{d\phi}{2\pi}\cos^2 2\phi 
 \left[ \log\left(\frac{1}{\sqrt{e}}\frac{2\omega_c}{\Gamma+\pi T +\sqrt{(\Gamma+\pi T)^2+\Delta_\phi^2}}\right) \right.  
+ \left. \frac{\Gamma+\pi T}{2\sqrt{(\Gamma+\pi T)^2+\Delta_\phi^2}} \right].
\label{eq:p04}
\end{equation}
By examining Eq.~(\ref{eq:p04}), we can clearly
see that the logarithmic divergence of the first term is cut-off by both $\Gamma$ and $\Delta_0$, so
a solution is no longer guaranteed.
We can estimate the critical values of the inverse lifetime and pseudogap to kill superconductivity at $T$=0
for limiting cases. In the clean limit with $\Gamma=0$, $\Delta_{cri}=2\pi e^{-(\gamma +1)}$T$_{c0}$ where $\gamma$ is Euler's constant
and T$_{c0}$ is T$_c$ for $\Delta_0, \Gamma = 0$.
With no pseudogap, we find a critical inverse lifetime $\Gamma_{cri}$
of $\pi e^{-\gamma} T_{c0}/2$ (Abrikosov-Gor'kov \cite{AG}). Fig.~3 shows 
the numerically evaluated left hand side of Eq.~(\ref{eq:Tcweak}) (denoted as $\lambda^{wc}_{max}$) as a function of temperature
for various $\Delta_0$, with the variation of $T_{c}$ with $\Delta_0$ or $\Gamma$ shown in the inset.  One clearly sees the logarithmic
divergence is cut-off as $\Delta_0$ increases, leading to a maximum in $\lambda$ at a particular temperature.  Once this maximum falls
below unity, no superconducting solution exists.

In order to show that our findings are general and not limited to weak coupling assumptions,
we consider a calculation based on a $V$ derived from a phenomenological form for $\chi$ \cite{MMP,Abanov}:
\begin{equation}
 V(k,\Omega) = \frac{3}{2} g_{sf}^{2} \frac{\chi_{\bf{Q}}}{\xi^{-2}_{AF}+2+\cos k_x + \cos k_y -i \frac{\Omega}{\Omega_{sf}}}
\end{equation}
where $g_{sf}$ is the coupling between fermions and spin 
fluctuations, $\xi_{AF}$ is the antiferromagnetic coherence length, $\Omega_{sf}$ is the
characteristic spin fluctuation energy scale, and $\chi_{\bf{Q}}$ is the static susceptibility
at the commensurate vector $\bf{Q}=(\pi,\pi)$.  For illustrative purposes, we take
$g_{sf}^{2}\chi_{\bf{Q}}=0.27$ eV, $\xi_{AF}=10$, $\Omega_{sf}=0.4$ eV with a cutoff energy
for Im $\chi$ of 0.4 eV, though we have studied a variety of parameter sets
(particularly variation of $\xi_{AF}$).
In general, these parameters are temperature dependent, but for
simplicity we ignore this.
We use the same model $G$ from above which was used to
study the weak coupling limit. Fig.~4a 
shows  the variation of $T_c$ with the pseudogap 
for different values of $\Gamma$. As in the weak coupling case, $\Delta_0$ and $\Gamma$
suppress T$_c$.  As expected, the size of $\Delta_0$ needed to destroy superconductivity is
of order T$_{c0}$.
The behavior of $\lambda$ with temperature is similar to the weak coupling case, as illustrated in
 Fig.~4b.   Again, a solution fails to appear once the temperature maximum of $\lambda$
 falls below unity.   The same behavior was found in the Fermi surface restricted results presented in Fig.~1.
 In turn, use of our phenomenological $G$ and $\chi$ in the full Brillouin zone formalism leads to similar
 behavior to Fig.~1 as well, with weakly temperature dependent $\lambda$ having values
 much less than unity (see Supplementary Information).

Over much of the doping-temperature phase diagram of the cuprates, ARPES reveals strongly lifetime broadened
features with a large pseudogap above T$_c$.  Despite this, T$_c$ is large except under extreme
underdoping conditions.  The work presented above indicates that for such a large pseudogap,
there should be no superconducting solution.  In our phenomenological studies, this can be mitigated
somewhat by using model Green's functions \cite{Norm07} which have Fermi surfaces in the pseudogap phase (as occurs
with charge ordering, spin ordering, or more phenomenological considerations like those of Yang,
Rice and Zhang \cite{YRZ}).  On the other hand, the fact that we find this same behavior using experimental Green's
functions indicates that this is a general issue, not specific to any particular model.

There is a way out of this dilemma.  If the pseudogap were due to pairing,
then all of the above conclusions are invalidated.  In this case, the mean field T$_c$ would actually
be the temperature at which $\Delta_0$ becomes non-zero, with the true T$_c$ suppressed from this due to fluctuations.
In a preformed pairs picture, T$_c$ would be controlled by the phase stiffness of the pairs \cite{Emery},
whereas in RVB theory, it would be controlled by the coherence temperature of the doped holes \cite{Wen}.
Regardless, our results are in strong support for such models.  ARPES \cite{Amit,Yang} and tunneling (STM) \cite{Kohsaka,Alldredge} are
consistent with a pairing pseudogap, since the observed spectra associated with the antinodal region of the zone
have a minimum at zero bias as would be expected if the gap were due to pairing (local or otherwise).  This does not mean that charge and/or
spin ordering does not occur in the pseudogap phase, it is just that our results are consistent with
these phenomena not being responsible for the pseudogap itself.

\linespread{1}
\section*{Acknowledgments}
The authors thank Doug Scalapino for suggesting this work, and he and Andrey
Chubukov for several helpful discussions.
This work was supported by the US DOE, Office of Science, under
contract DE-AC02-06CH11357 and by the Center for Emergent
Superconductivity, an Energy Frontier Research Center funded by the
US DOE, Basic Energy Sciences, under Award No.~DE-AC0298CH1088.

\begin{figure}
\includegraphics[width=0.8\columnwidth]{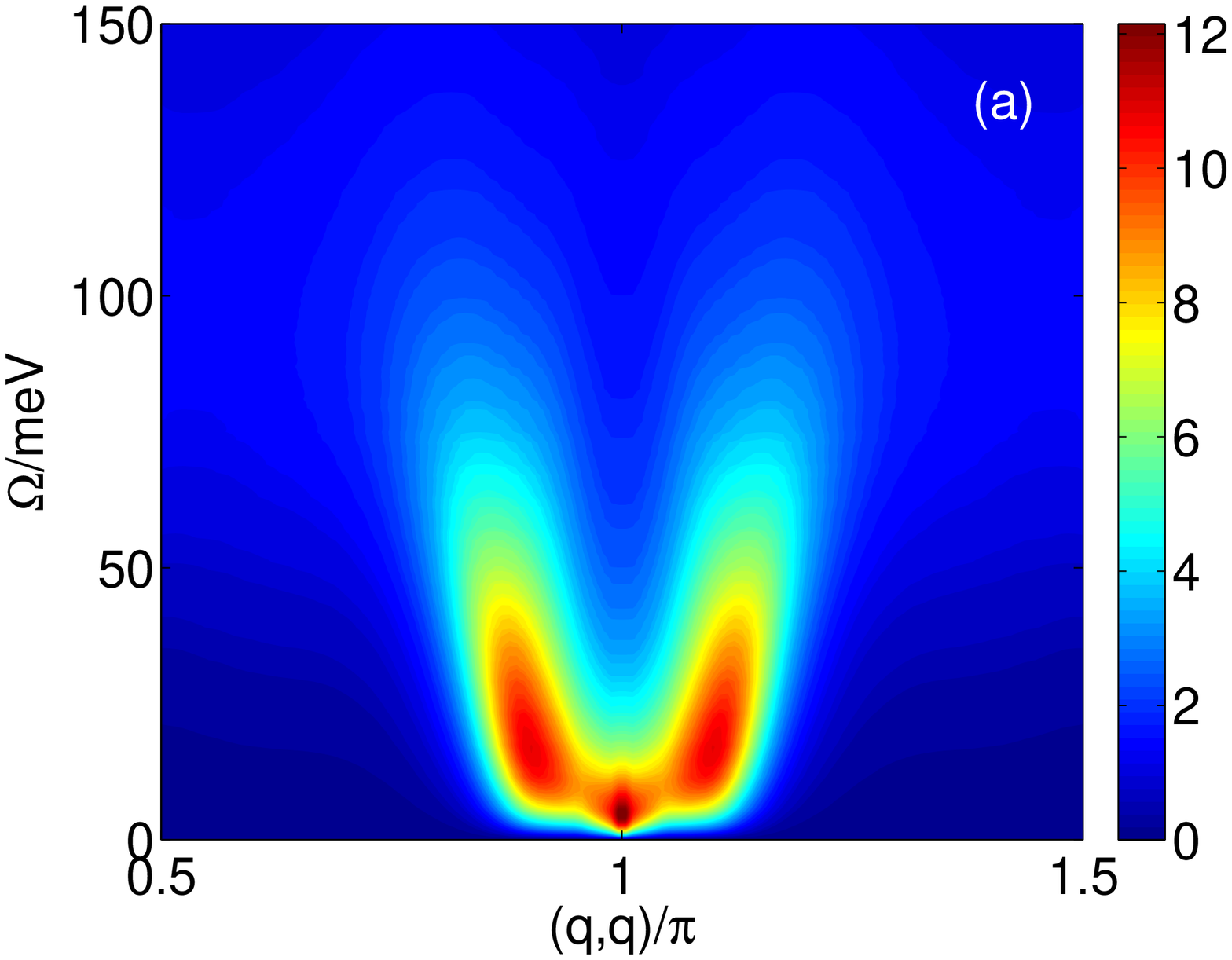}
\includegraphics[width=0.8\columnwidth]{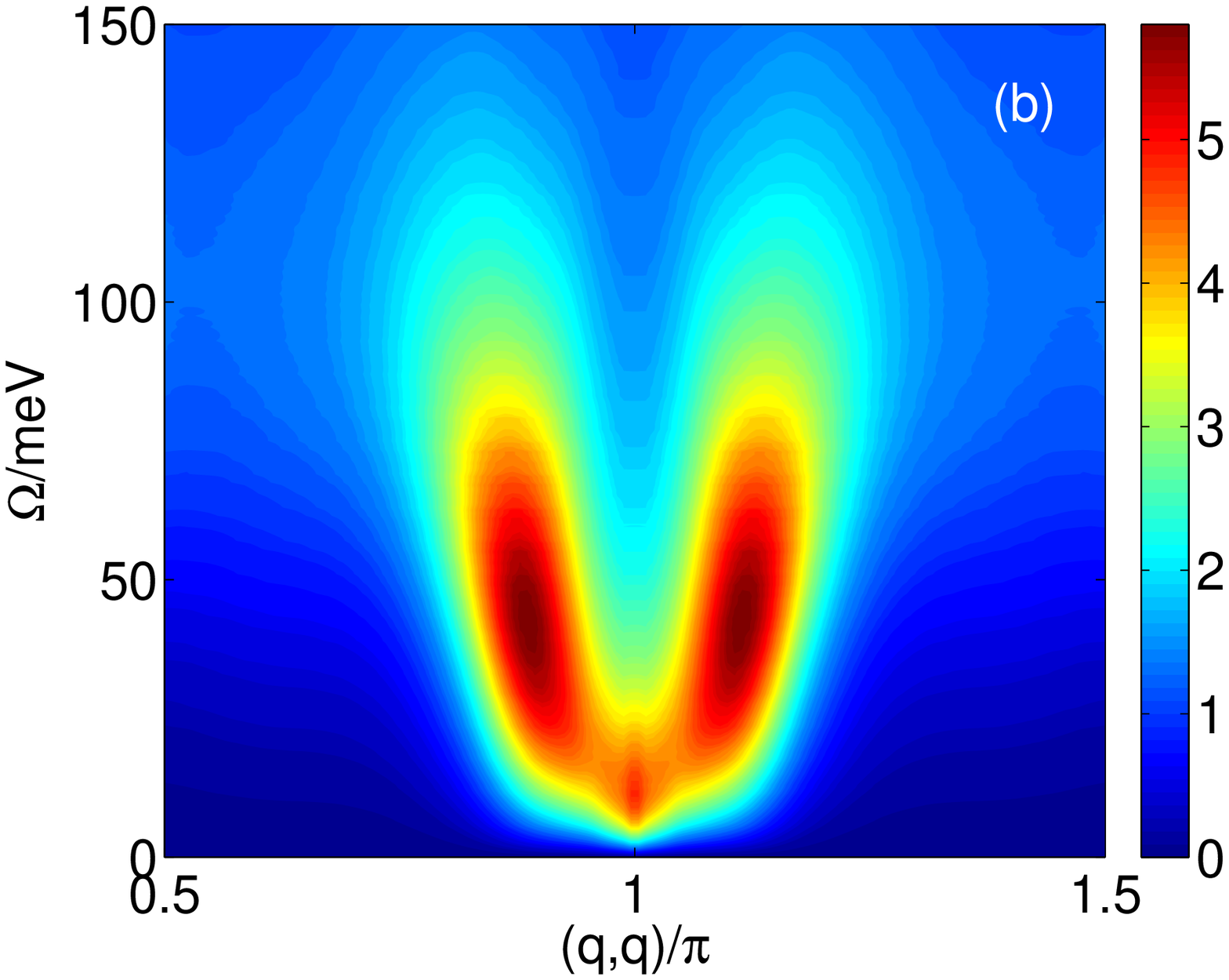}
\caption{Imaginary part
of the susceptibility $\chi^{\prime\prime}$ constructed from experimental
Green's functions versus energy along
the nodal direction for (a) $U$=860 meV and (b) 800 meV.}
\label{fig:chi_rpa}
\end{figure}

\begin{figure}
\includegraphics[width=0.9\linewidth]{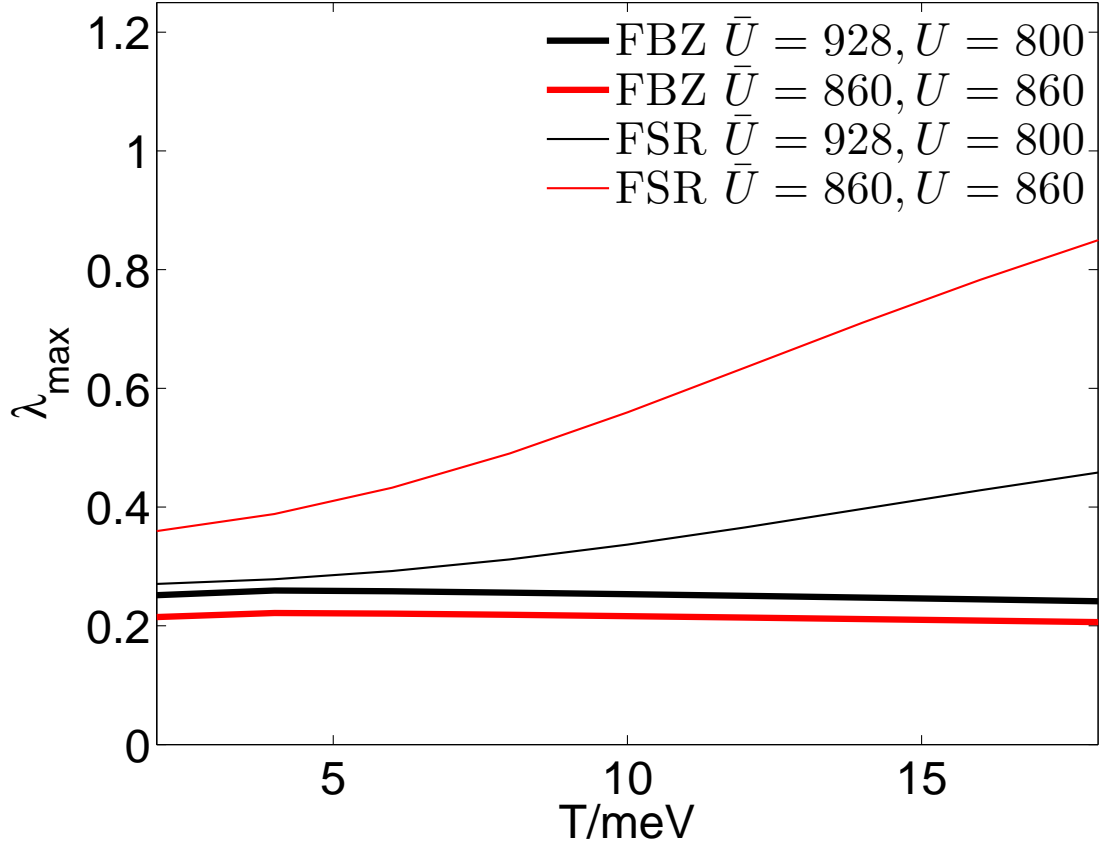}
\caption{The leading eigenvalue $\lambda_{max}$ as a function of temperature, using experimental
 spectral functions. FSR (thin curves) represent Fermi surface
 restricted calculations, FBZ (thick curves) represent eigenvalues obtained from Eq.~(\ref{eq:gap_eq}) (full Brillouin zone).
 The interaction parameters are indicated in units of meV.}
\label{fig:eigenvalue}
\end{figure}

\begin{figure}
\includegraphics[width=0.9\linewidth]{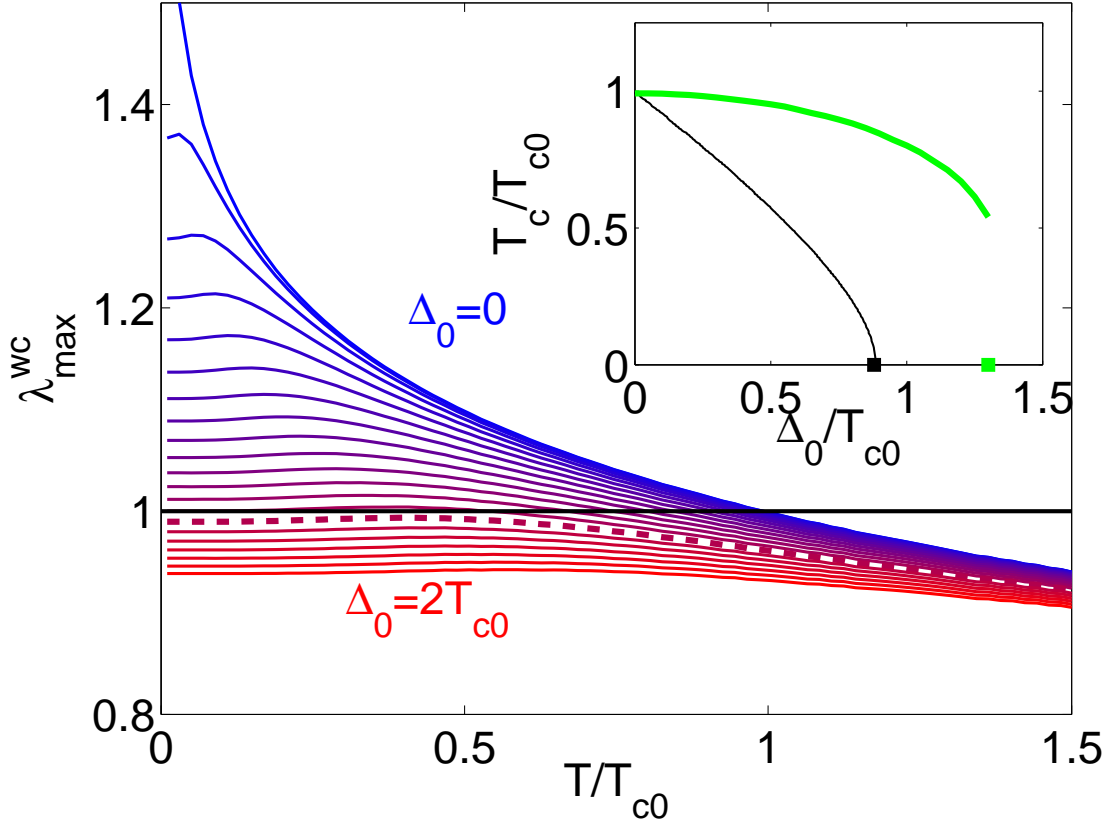}
\caption{$\lambda^{wc}_{max}$ for a
d-wave superconductor with
a d-wave pseudogap as a function of temperature plotted for
various values of the pseudogap. This quantity (left hand side of Eq.~(\ref{eq:Tcweak}), with $\Gamma=0$)  is the weak coupling
analog of $\lambda_{max}$. All energies
are normalized to the mean field transition temperature T$_{c0}$ for $\Gamma, \Delta_0 = 0$.
For large enough $\Delta_0$, a solution does not exist (dashed curve and ones below it). 
Inset: T$_c$ as a function of the pseudogap. The curve abruptly terminates when the maximum
in $\lambda$ as a function of $T$ goes below unity.
The thin curve shows
T$_c$ as a function of the inverse lifetime $\Gamma$.  In this case,
the x axis should be read as $\Gamma/T_{c0}$.
The filled boxes are the analytic estimates at $T$=0.
}
\label{fig:tc_wk}
\end{figure}

\begin{figure}
\includegraphics[width=0.8\columnwidth]{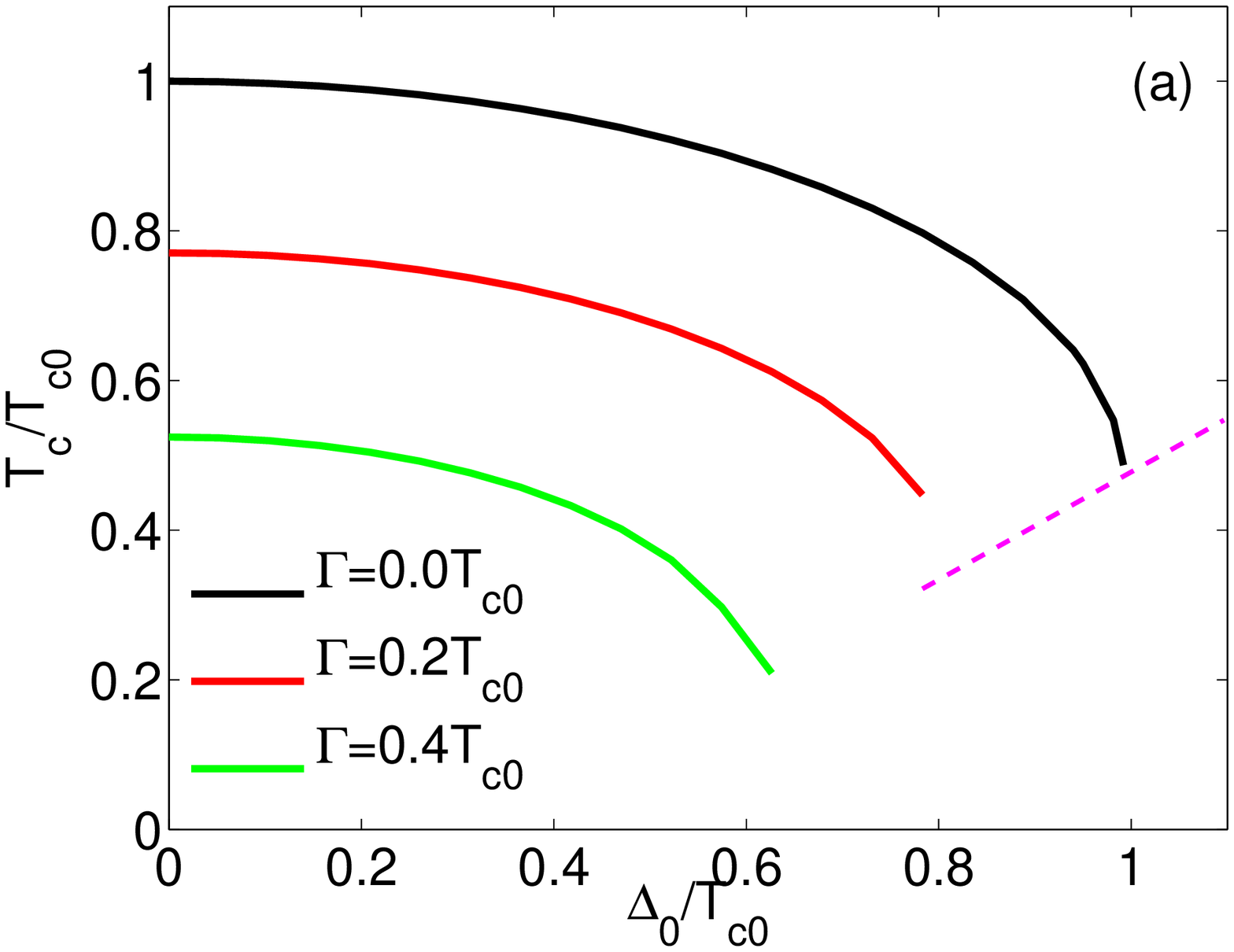}
\includegraphics[width=0.8\columnwidth]{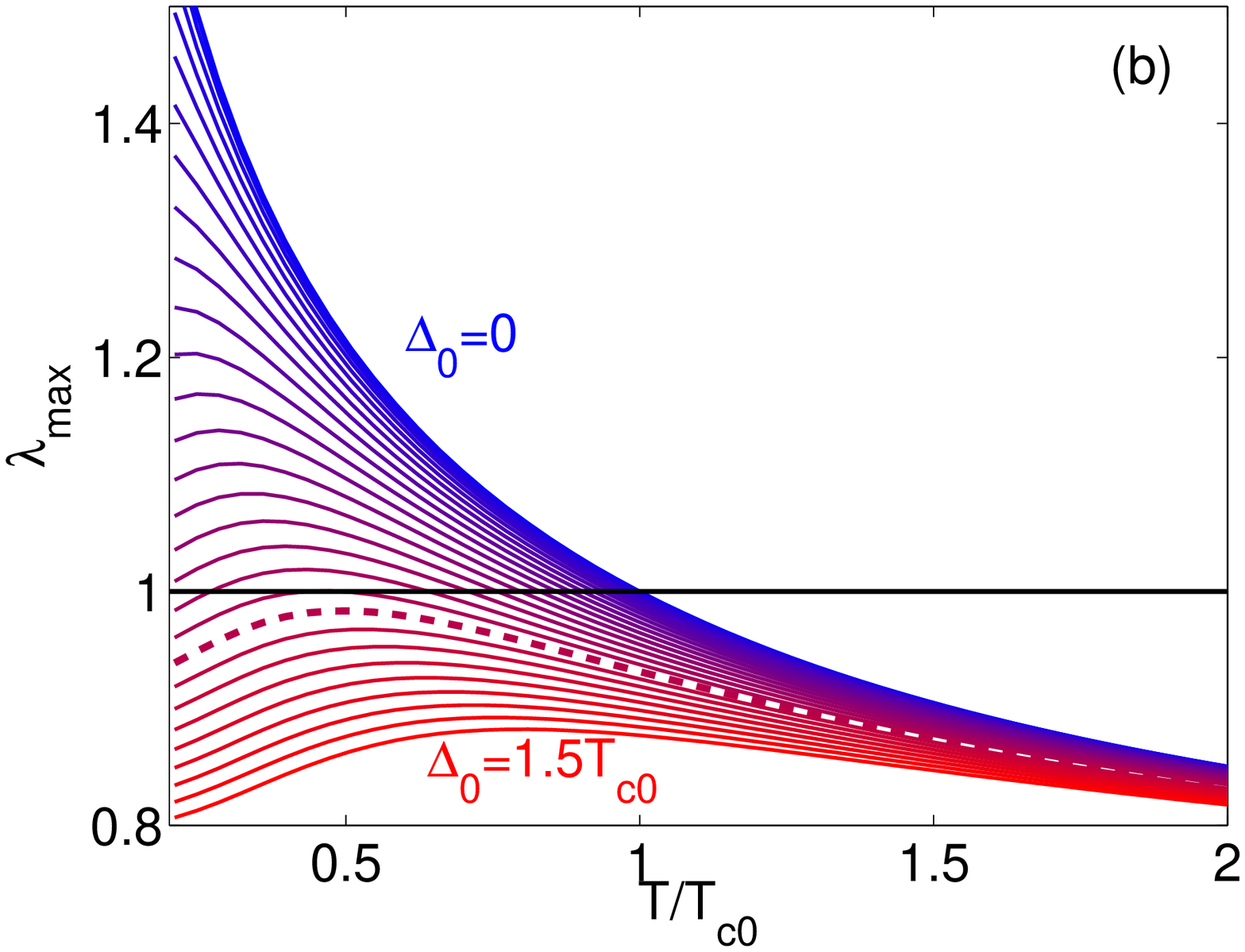}
\caption{Transition temperature as a function of the pseudogap $\Delta_0$ for
various values of the inverse lifetime $\Gamma$ (a). The energy scales are normalized
to the value of T$_{c0}$ (4.8 meV) for $\Delta_0, \Gamma = 0$.  The dashed curve is
the variation of the temperature maximum in $\lambda$ for the $\Gamma=0$ case.  Once the T$_c$
curve intersects this, no solution exists.  This is evident from (b), where the temperature
dependence of $\lambda_{max}$ is plotted for various $\Delta_0$. For the dashed curve and below,
no solution exists.}
\label{fig:tc}
\end{figure}

\clearpage
\linespread{1}
\section*{Supplementary material}
\linespread{1}
\subsection{Pairing equations}
For the full Brillouin zone calculations, we use a 64 by 64 point grid in the first
Brillouin zone and a Matsubara cut-off of 40.  For the Fermi surface restricted calculations, we use an angular step of 1 
degree on the Fermi surface, with a Matsubara cut-off of 100.  Convergence was tested for both the $k$ point sum and the Matsubara
cut-off.  The tight binding fit used for $\xi_k$ was that of Kaminski {\it et al} \cite{Adam05}.

Technically, we should be solving two coupled equations, one for $\Phi$ and one for $Z$.  But since we are equating
$G$ to the experimental Green's function from ARPES, we do not solve the $Z$ equation.  We note that the popular
`trick' of reducing these two equations to a single master equation in the Fermi surface restricted case does not work
in the presence of a pseudogap.  That is, the pseudogap can be represented in the functional form \cite{Norm07s}
\begin{equation}
\Sigma_{PG} = \frac{\Delta_k^2}{\omega - X_k}
\end{equation}
which can be easily generalized to include broadening.  Here, $X_k$ is $-\xi_k$ for the pairing case, and $\xi_{k+Q}$
for density wave ordering at $Q$.  The important point is because of the strong dependence of $X_k$ on $k_{\perp}$,
one cannot collapse to a simple master equation commonly used in Eliashberg calculations \cite{Abanov08}.
We admit, though, that because we do not solve the $Z$ equation (except to estimate $\bar{U}$), we could be overestimating
pair breaking effects.

\subsection{Temperature dependent models}
\begin{figure}
\includegraphics[width=0.9\linewidth]{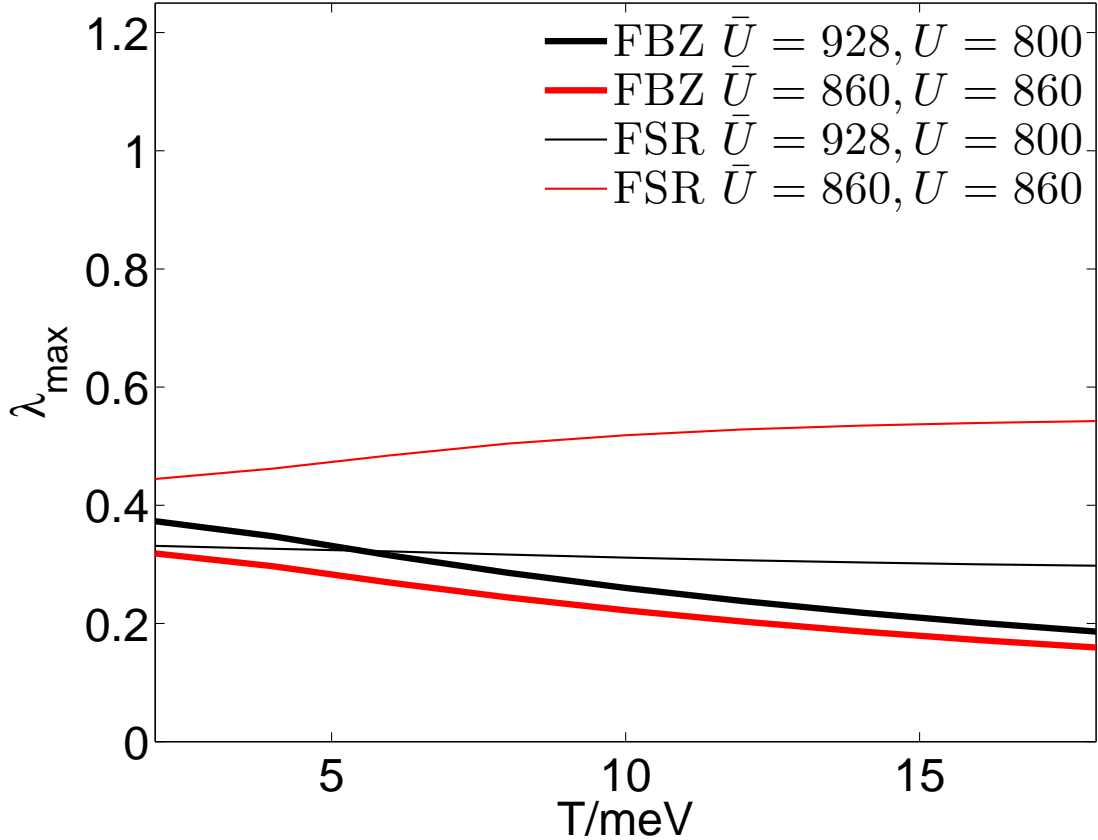}
\caption{The leading eigenvalue $\lambda_{max}$ as a function of temperature, as in Fig.~2, but using for $GG$ the model
 spectral function of Eq.~(\ref{eq:model_akw}) with a $T$ dependent $\Gamma$.
The interaction parameters are indicated in units of meV.}
\label{fig:eigenvalue_model}
\end{figure}

To mimic the temperature dependence of the spectral function, we can allow $\Gamma$  and $\Delta_0$ to be $T$ dependent in
\begin{equation}
 A(k,\omega)=-\frac{1}{\pi}Im\left[\frac{\omega+i\Gamma+\xi_k}{\left(\omega+i\Gamma\right)^2-\xi_{k}^{2}-\Delta_{k}^{2}}\right].
 \label{eq:model_akw}
\end{equation}
Based on the ARPES data, we take $\Delta_0$ to be temperature independent.  For the Fermi surface restricted calculations,
it has the form $\Delta_0 \cos(2\phi)$ with a $\Delta_0$ of 50 meV.  For the full Brillouin zone calculations, it has the form
$\Delta_0 (\cos(k_x)-\cos(k_y))/2$ with a $\Delta_0$ of 54 meV.  In both cases, we again use $\chi$ constructed from the
experimental Green's functions, and so the $A$ in Eq.~(\ref{eq:model_akw}) is just used in the $GG$ part of the pair vertex.
With a temperature independent $\Gamma$=50 meV, 
we find similar behavior in the pairing equation as when we use
the experimental spectral functions.
Next we consider a more realistic model where $\Gamma$ behaves linearly with $T$ as in marginal Fermi liquid theory
(here, we use a coefficient in front of T of 5.58 to reproduce the pseudogap temperature T$^*$ of 180 K for this sample,
since this form for $G$ will becomes gapless at the anti-node \cite{Norm07s} once $\Gamma = \sqrt{3} \Delta_0$).
Again, the behavior of $\lambda_{max}$ with T is similar to before, as shown in Fig. \ref{fig:eigenvalue_model}.

\end{document}